\documentclass{ws-procs9x6}
\begin{document}
\title{Deeply Virtual Compton Scattering with CLAS}
\author{F.X. Girod$^*$~(for the CLAS collaboration)}
\address{Hall-B, Jefferson Laboratory\\
Newport News, VA 23606, USA\\
$^*$E-mail: fxgirod@jlab.org}
\begin{abstract}
The beam spin asymmetries of the reaction $\vec{e}p\rightarrow ep\gamma$ in the Bjorken regime
were measured over a wide kinematical domain using the CLAS detector and a new lead-tungstate calorimeter.
Through the interference of the Bethe-Heitler process with Deeply Virtual Compton Scattering,
those asymmetries provide constraints for the nucleon Generalized Parton Distributions models.
The observed shapes are in agreement with twist-2 dominance predictions.
\end{abstract}
\keywords{CLAS; DVCS; GPD.}
\bodymatter
\section{Introduction}
The introduction of parton Wigner distributions in the nucleon~\cite{Ji03}
provided us with a rigorous landscape of the different approaches to the hadron structure,
linking together all possible hadronic observables.
There is unfortunately as of today no known experimental process to access them.
Projecting out light-cone energy and transverse momenta, a Fourier transform leads 
to Generalized Parton Distributions~(GPDs).
Those encompass Form Factors and Parton Distribution Functions, 
as well as a wealth of new physical informations. For instance,
the Energy-Momentum tensor of partons in the nucleon can be parameterized in terms of GPDs,
providing distributions of masses, angular momenta, or forces inside the nucleon~\cite{Schw07}.

In the Bjorken limit, the amplitude for Deeply Virtual Compton Scattering (DVCS) $\gamma^*p\rightarrow\gamma p$ 
factorizes~(as illustrated in figure~\ref{fig:handbag}), and the scattering occurs at the quark level.
This is considered to be the cleanest process to access the GPDs.
First observations~\cite{H1-01} are in agreement with twist-2 dominance in the GPDs formalism. 
A first dedicated experiment~\cite{Mun06} demonstrates that scaling is achieved at moderate values of $Q^2$, as expected both from
theoretical considerations and the successfull description of DIS down to $Q^{2}\approx 1$ GeV$^{2}$.

In the reaction $\vec{e}p\rightarrow ep\gamma$ , the beam-spin asymmetry of the reaction results from 
the interference between DVCS and Bethe-Heitler processes, for which the outgoing photon was emitted by the electron.
This asymmetry~\footnote{The arrows correspond to the beam helicity.}~:
$$\text{A}=\frac{\text{d}^4\overrightarrow{\sigma}-\text{d}^4\overleftarrow{\sigma}}
{\text{d}^4\overrightarrow{\sigma}+\text{d}^4\overleftarrow{\sigma}}\stackrel{\text{twist-2}}{\approx}\frac{\alpha\sin\phi}{1+\beta\cos\phi}$$
\noindent has been measured as a function of $Q^2$, $x_B$, 
the momentum transfer to the proton $t$, and the angle\footnote{We use the Trento convention.} $\phi$
between the leptonic and hadronic planes.
Harmonic decompositions~\cite{DGPR97} have been proposed for the cross-sections.
The cross-section difference $\text{d}^4\overrightarrow{\sigma}-\text{d}^4\overleftarrow{\sigma}$ is proportional to the
imaginary part of the DVCS amplitude, and sensitive to a specific combination of GPDs $H$, $\tilde{H}$ and $E$ at $x=\pm\xi$. 

\section{Experimental context}
Using JLab continuous electron beam at energy $5.8$ GeV, with average $80$\% polarisation and intensity $25$ nA 
on a 2.5~cm-long liquid hydrogen target, we collected $45$ fb$^{-1}$.
A new lead-tungstate calorimeter was added to CLAS to detect photons between 4.5 and 15$^\circ$.
Avalanche photodiodes collect the light out of the 424 tapered crystals 16~cm long, with average section 2.1~cm$^2$,
each monitored with laser light.
Resolutions achieved in energy are better than 5\% at 1~GeV, in angle 3 to 4~mrad, and in time less than 1~ns.
A superconducting solenoidal magnet actively shields the background of M{\o}ller electrons at low angles and energies.
\section{Analysis and results}
We selected events with one and only one electron (trigger), proton and photon.
The electron identification uses mainly deposited energy in CLAS calorimeters compared to momentum measured by drift chambers,
and \v{C}erenkov counters.
The protons were separated from positive pions through time-of-flight and track length compared to momentum.
Photon detection threshold was set at 150 MeV.

The mere selection of the 3-particle final state results in clear exclusive peaks in the momentum and energy balance spectra.
In addition, we required the missing transverse momentum less than 90 MeV/c, 
the detected photon within 1.2$^\circ$ from the predicted one,
and the angle between the $\gamma^*p$ and $\gamma p$ planes smaller than 1.5$^\circ$.
The neutral pion electroproduction $ep\rightarrow ep\pi^0$ can still contaminate our $ep\rightarrow ep\gamma X$ sample, 
in the case were the $\pi^0$ decay was asymmetric with one of the photon below the energy threshold, or outside the calorimeters' acceptance.
This contamination was estimated in a model-independent way from simulation and subtracted in each elementary bin and for each helicity state.

\begin{figure}[h]
\begin{minipage}[b]{.43\textwidth}
\psfig{file=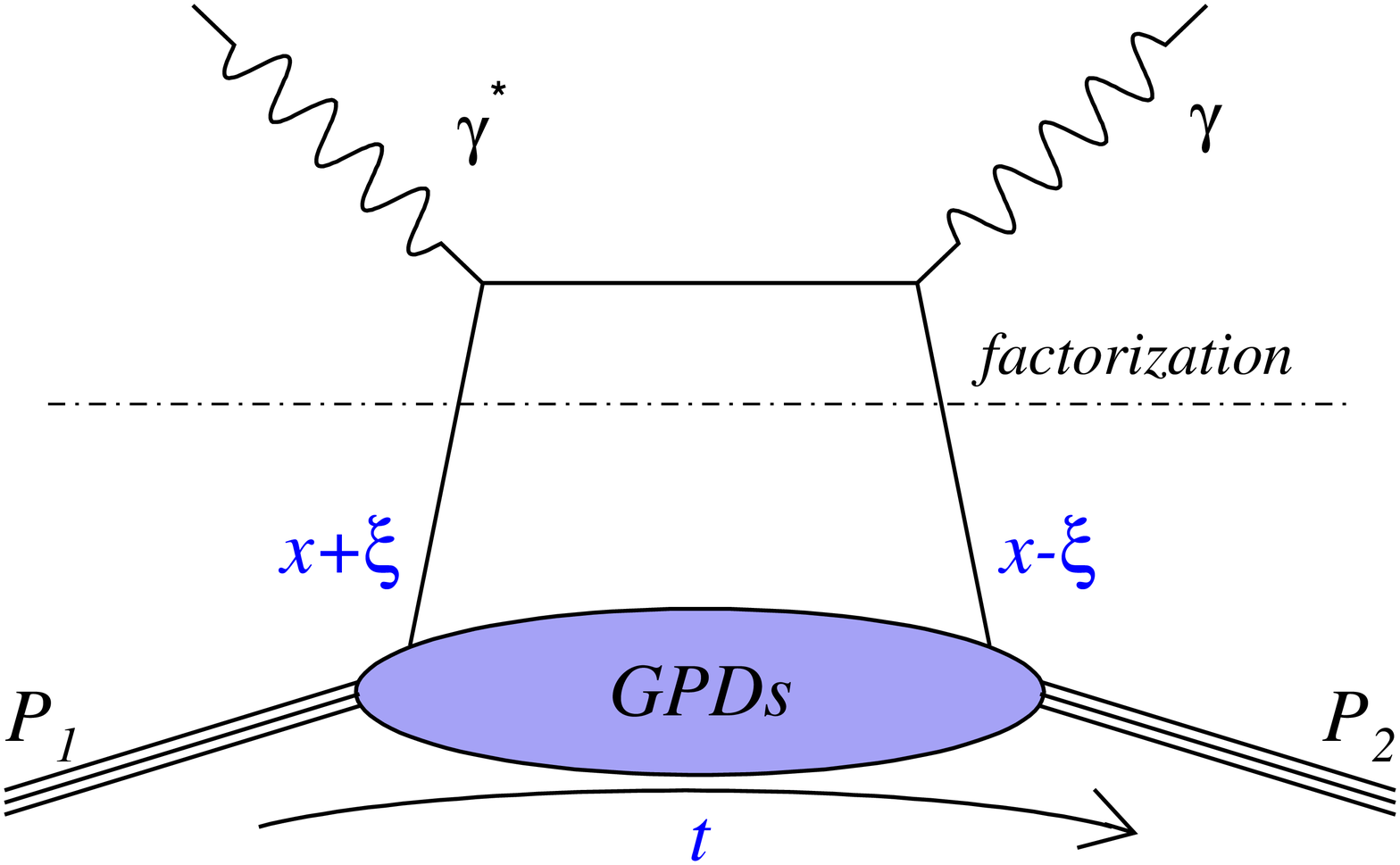,width=2.0in}
\caption{Handbag diagram for DVCS. $t$ is the squared 4-momentum transfer to the proton,
$x$ is the average longitudinal momentum fraction, in terms of $(p+p')/2$,  of the active quark in the initial and final states, and 
$2\xi\approx x_B/(2-x_B)$ parameterizes the difference.}
\label{fig:handbag}
\end{minipage}
\begin{minipage}[t]{.53\textwidth}
\psfig{file=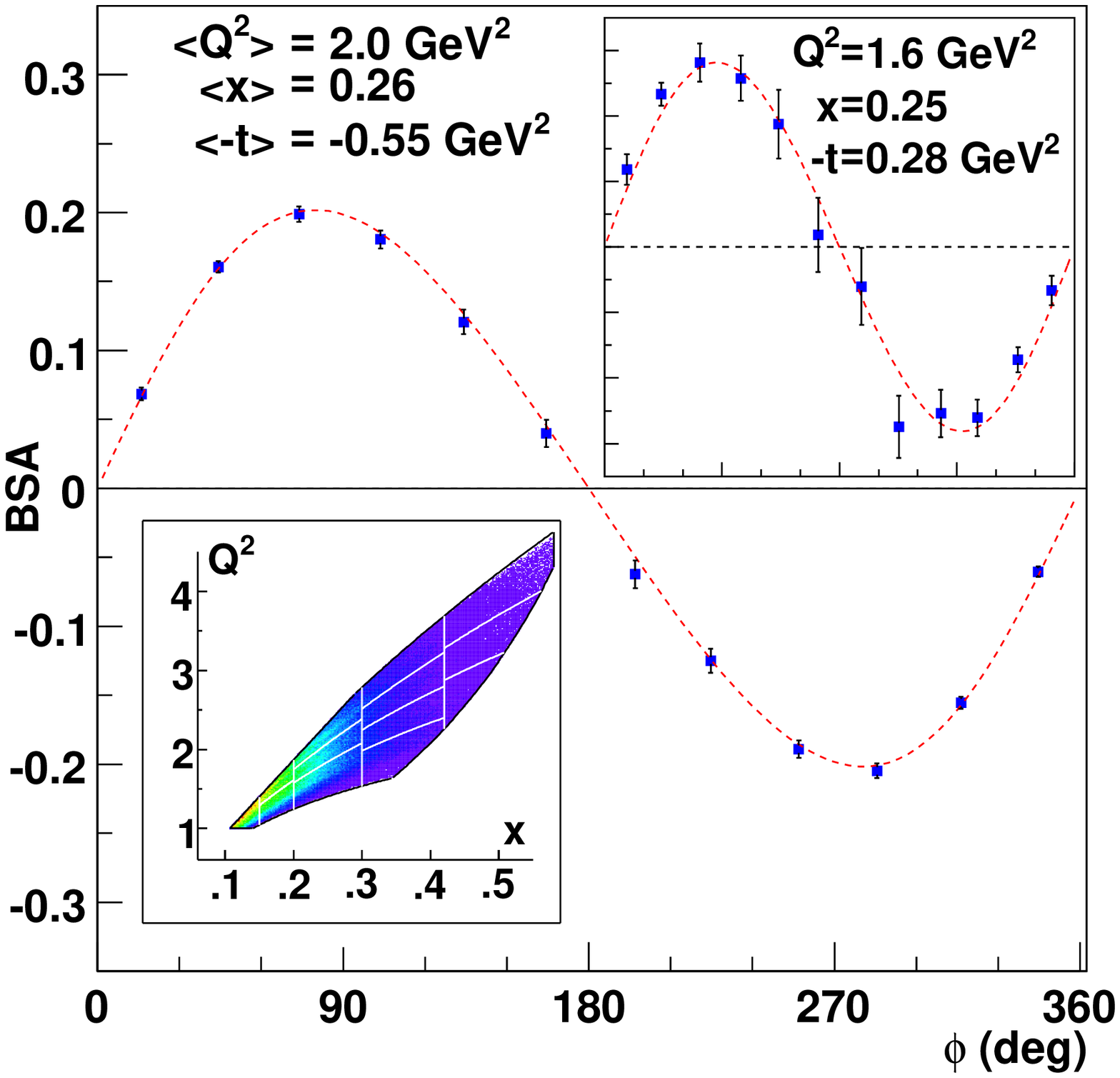,width=2.5in}
\vspace{-0.45in}
\caption{Integrated result for the asymmetry.
Bottom-left insert~: $(x_B,Q^2)$ kinematical coverage.
Top-right insert~: one elementary bin.}
\label{fig:BSAint}
\end{minipage}
\begin{minipage}[b]{.95\textwidth}
\psfig{file=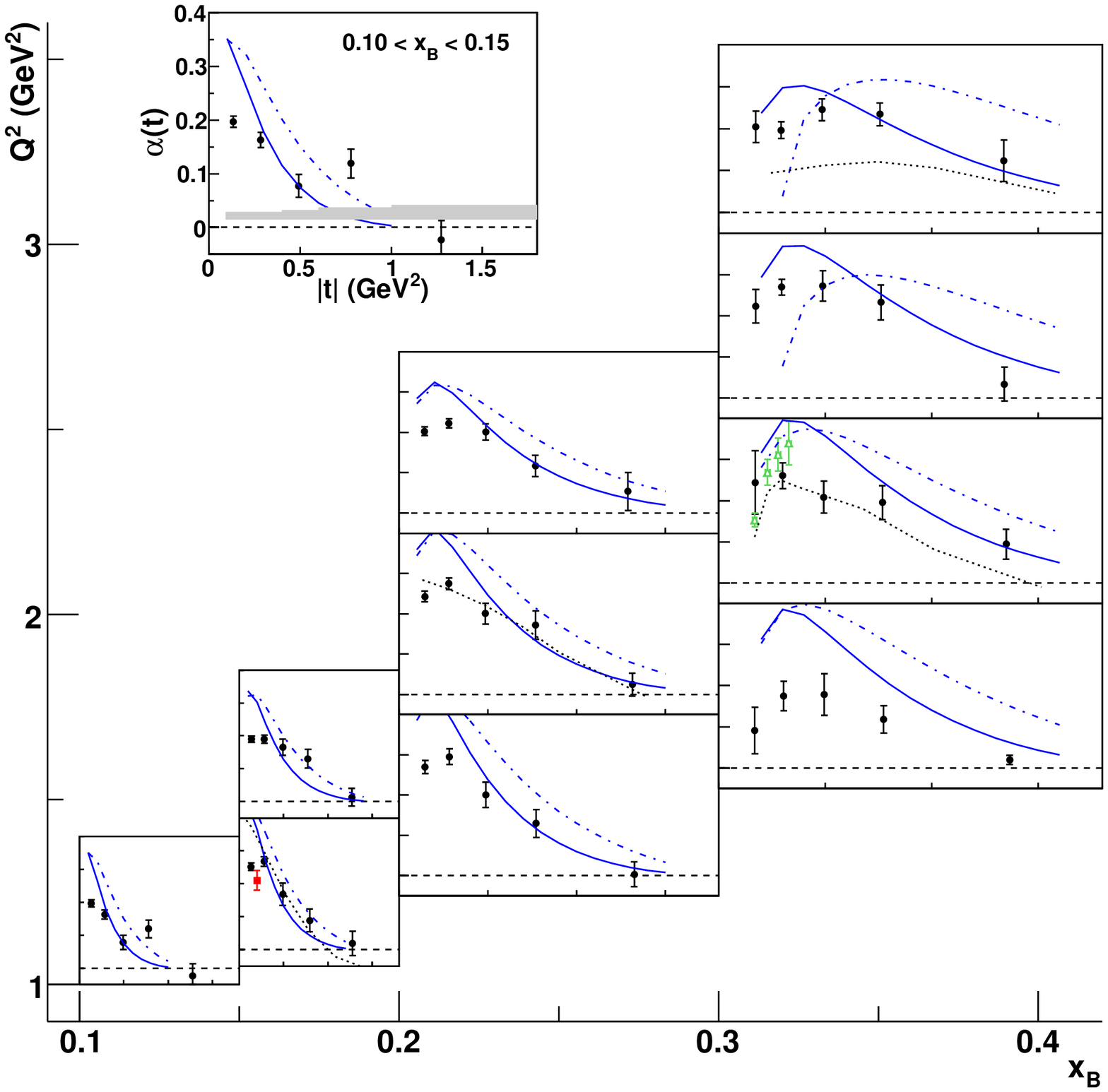,width=4.5in}
\caption{BSA at $\phi=\pi/2$ estimated by the $\alpha$ parameter, as a function of $-t$ (see text for details).}
\label{fig:BSA_vs_t}
\end{minipage}
\end{figure}

The integrated asymmetry is displayed in figure~\ref{fig:BSAint}, together with the kinematical coverage in the $(x_B,Q^2)$ plane, 
and one single $(x_B,Q^2,t)$ elementary bin fitted with expected shape.
The parameter $\alpha$ estimates the asymmetry at $\phi=\pi/2$ and is shown in (black) circles as a function of $-t$ 
in the $(x_B,Q^2)$ plane in figure~\ref{fig:BSA_vs_t}.
Also appearing are results from CLAS first observation in a (red) square, and Hall-A in (green) triangles.
A GPD semi-phenomenological model~\cite{VGG} appearing in (blue) solid (twist-2) 
and dashed-dotted (twist-3\footnote{Kinematical twist-3 only, in the Wandzura-Wilczek approximation.}) 
lines in this plot reproduces reasonably the data.
A Regge-inspired calculation~\cite{JML} in dotted lines proves fair agreement in some bins.
A possible duality between those two descriptions, in terms of Regge-exchanges and GPD formalism, remains to be investigated. 

The succesful operation of CLAS and its new calorimeter allowed measurement of the most extensive set of DVCS data, 
providing strong constraints for GPDs model and future extraction.

\end{document}